\documentstyle[pre,aps,epsfig,amsmath,amssymb,color]{revtex}
\begin{document}
\draft  
\title{Magnetic order in the Ising model with parallel dynamics}
\author{Emilio N.M.\ Cirillo}
\address{Dipartimento Me.\ Mo.\ Mat.,  Universit\`a degli Studi di Roma
``La Sapienza," via A.\ Scarpa 16, 00161 Roma, Italy}
\author{Francesca R.\ Nardi}
\address{Eurandom, PO BOX 513, 5600MB, Eindhoven, Netherlands}
\author{Antonio D.\ Polosa}
\address{Department of Physics, University of Helsinki,
PO BOX 64, FIN-00014, Finland}
\date{\today}
\maketitle
\begin{abstract}
It is discussed how the equilibrium properties of the Ising model
are described by an Hamiltonian with an antiferromagnetic low
temperature behavior if only an heat bath dynamics, with the
characteristics of a Probabilistic Cellular Automaton, is assumed
to determine the temporal evolution of the system.
\end{abstract}


\definecolor{light}{gray}{.8}
\newcommand{\commento}[1]{
$\phantom .$
\bigskip
\par\noindent
\colorbox{light}{\begin{minipage}{13cm}#1\end{minipage}}
\bigskip
\par\noindent
}

\section{Introduction}
\label{s:intro}
\par\noindent
In this note we discuss the equilibrium properties of
Probabilistic Cellular Automata (PCA) reversible with respect to a
Gibbs measure derived by a suitable Hamiltonian. A PCA
\cite{[St],[KV],[V]} is a
lattice model with discrete variables which are subject to a
probabilistic {\it simultaneous} updating in discrete time steps:
all configurations are accessible in a single updating.

PCA arise as an extended definition of Deterministic Cellular
Automata in which the updating follows a set of deterministic
local rules. The huge number of possible deterministic (or
probabilistic) rules makes the topic of Cellular Automata
overwhelmingly abundant. One of the most famous Cellular Automata
systems is the Conway's  ``Game of life''\cite{croiz}; in spite of
the very simple deterministic majority rule assigned, the system,
which is a kind of spin lattice, presents an extremely rich and
complex evolution pattern.

PCA have been studied in a wide variety of contexts, ranging from
biology to the theory of automation, while their role in
statistical mechanics has been thoroughly investigated only
relatively recently in \cite{[R],[D],[LMS1],[LMS2],[BCLS]}.
The particular family
of automata we study is obtained by implementing in parallel
fashion the heat bath dynamics \cite{[D]}. In other terms, we define a rule
for the transition probabilities such that all single spins of a
lattice are updated simultaneously with heat bath rates. This
amounts to define a Markov chain for the evolution of the spin
system, having the characteristics of a PCA.

We observe that the way of implementing the heat bath dynamics
reflects into a qualitative modification of the equilibrium
properties of the model. In particular, an Ising-like
ferromagnetic Hamiltonian with two body interactions, defines a
PCA reversible with respect to a Gibbs measure determined by an
Hamiltonian  allowing a low temperature antiferromagnetic phase.
This behavior is absent if a  serial dynamics is implemented, for
which at most one spin of the system is updated at any time.

The paper is organized as follows. In Section \ref{s:rev} we define the PCA
under consideration and the spin model Hamiltonian, $H$,
determining the heat bath single spin rates.
In Section \ref{s:ising}
the structure and the low temperature antiferromagnetic properties
of $H^\prime$ are discussed. There follows a discussion of its
equilibrium properties and ground states. Finally, in Section \ref{s:next}, it
is considered the case in which  the heat bath spin rates are
determined  by a (two body) next-to-nearest neighbor interaction.
The dynamical generation of antiferromagnetic couplings is
reviewed in the conclusive remarks.

\section{Coupling proliferation in reversible PCA's}
\label{s:rev}
\par\noindent
Let $\Lambda$ be a finite two--dimensional square lattice and
$|\Lambda|$ its cardinality. For each
$x=(x_1,x_2),y=(y_1,y_2)\in\Lambda$ we denote by
$|x-y|$ the Euclidean distance
on the lattice. Let $\sigma(x)\in\{-1,+1\}$ a spin variable
associated to the site $x\in\Lambda$; the space $\{1,-1\}^\Lambda$
of configurations is denoted by $\mathcal{S}$.

Let us consider a generic Hamiltonian $H:\sigma\in{\mathcal S}\to
H(\sigma)\in\mathbb{R}$ and the corresponding equilibrium Gibbs
measure
\begin{equation}
\label{gibbs}
\mu(\sigma)= \frac{e^{-\beta H(\sigma)}}
                  {\sum_{\eta\in\mathcal{S}}e^{-\beta H(\eta)}} \;\;\; ,
\end{equation}
with $\beta$ the inverse of the temperature. We now define the
heat bath single spin rates: given the site $x\in\Lambda$, we
consider the Gibbs equilibrium measure for $\sigma_x$ with respect
to a fixed configuration $\sigma$ on $\Lambda\setminus\{x\}$.
Letting $a\in\{-1,+1\}$, we have
\begin{equation}
p_x(a|\sigma) = \frac{\exp\left\{-\beta H
          (a,\sigma)\right\}
     }
     {\exp\left\{-\beta H
            (a,\sigma)\right\}
     +\exp\left\{-\beta H
            (-a,\sigma)\right\}
     } \;\;\; ,
\label{heat}
\end{equation}
where $(\pm a,\sigma)$ are the configurations equal to $\sigma$ on
$\Lambda\setminus\{x\}$ and to $\pm a$ on $x$. Note that the
normalization condition $p_x(a|\sigma)+p_x(-a|\sigma)=1$ is
trivially satisfied.

We can now implement the heat bath dynamics in a serial fashion,
namely we can consider the Markov chain $\sigma_t$ ($t$ being the
discrete time temporal variable), with transition probabilities
\begin{equation}
P(\sigma, \eta) =\left\{
\begin{array}{cl}
(1/|\Lambda|) p_x(\eta(x)|\sigma) & \;\;\;\textrm{if }\;\exists
x\in\Lambda\textrm{ such that }
  \sigma=\eta \;\;\;\textrm{on}\;\;\ \Lambda\setminus\{x\}
\\
0 &\;\;\;\textrm{ otherwise }
\end{array}
\right. \label{ser}
\end{equation}
for all $\sigma,\eta\in\mathcal{S}$. The transition probabilities
(\ref{ser}) are reversible with respect to the Gibbs measure
(\ref{gibbs}), i.e., the detailed balance condition is satisfied
or, equivalently, the equilibrium measure is the Gibbs measure
(\ref{gibbs}).

A different point of view can be taken \cite{[D]}: we define the
transition probabilities $P(\sigma, \eta)$ in such a way that all
the spins are {\it simultaneously} and independently updated, in a
parallel fashion, with the heat bath rates (\ref{heat}). Thus,
instead of (\ref{ser}) we consider the Markov chain $\sigma_t$
defined by
\begin{equation}
P(\sigma, \eta)
=\prod_{x\in\Lambda}p_x\left(\eta(x)|\sigma\right)\;\;\;
\forall\sigma,\eta\in\mathcal{S}\;\;\; . \label{par}
\end{equation}
This amounts to define a Probabilistic Cellular Automata (PCA).

In general the equilibrium properties of the Markov chain
(\ref{par}) are not trivial, for instance it is not obvious that
there exists a Gibbs measure such that the detailed balance
principle is satisfied.

Let us consider, now, the case of the two body interactions and
suppose that the Hamiltonian has the form:
\begin{equation}
H(\sigma)=
-\sum_{x,y\in\Lambda}J_{x,y}\sigma(x)\sigma(y)
-\sum_{x\in\Lambda}h_x\sigma(x) \;\;\; , \label{h2}
\end{equation}
where $J_{x,y}\in\mathbb{R}$ are the pair couplings between spins at
sites $x,y$, and $h_x\in\mathbb{R}$
is the external magnetic field acting on the spin at site $x$. For physical
reasons we suppose that the pair couplings are symmetric, namely
$J_{x,y}=J_{y,x}$ for all $x,y\in\Lambda$. The heat bath single
spin rates are given by
\begin{equation}
p_x(a|\sigma)=
  \frac{1} {1+\exp\left\{-\beta\left[H(-a,\sigma)-H(a,\sigma)\right]\right\}}
 =\frac{1}{2}\left[1+a\tanh\beta S_x(\sigma)\right] \;\;\; ,
\label{heat2}
\end{equation}
where
\begin{equation}
S_x(\sigma)= \sum_{y\in\Lambda\setminus\{x\}}J_{x,y}\sigma(y)
              +h_x \;\;\; ,
\label{derr}
\end{equation}
for any $\sigma\in\mathcal{S}$ and $x\in\Lambda$. It is easy to show
\cite{[D]} that the Probabilistic Cellular Automaton (\ref{par})
with single spin rates (\ref{heat2}) is reversible with respect to
the Gibbs measure $\mu'$ on $\mathcal{S}$ associated to the Hamiltonian
\begin{equation}
H'(\sigma)= -\beta \sum_{x\in\Lambda}h_x\sigma(x)
-       \sum_{x\in\Lambda}\log\cosh\left[\beta S_x(\sigma)\right] \;\;\; .
\label{ham'}
\end{equation}
In other words the detailed balance condition
\begin{equation}
P(\sigma,\eta)\exp\{-H'(\sigma)\}=
P(\eta,\sigma)\exp\{-H'(\eta)\} \;\;\; ,
\label{dett}
\end{equation}
is satisfied for any $\sigma,\eta\in\mathcal{S}$. This means that
$H^\prime(\sigma)$ is  the equilibrium Hamiltonian of a system
governed by $H(\sigma)$ and evolving with the law (\ref{par}).

The choice of the two body interaction in equation (\ref{h2}) is
strictly connected to the reversibility of the resulting
Probabilistic Cellular Automaton (\ref{par}) \cite{[KV],[V]}. For
example, consider the three body interaction Hamiltonian
$H(\sigma)=
\sum_{x,y,z\in\Lambda}J_{x,y,x}\sigma(x)\sigma(y)\sigma(z)$ with
the three body couplings $J_{x,y,z}$ symmetric with respect to
permutations of the indices and such that $J_{x,y,z}\not=0$ if and
only if $x\not=y\not=z\not=x$. Then the problem of showing the
reversibility of the parallel dynamics can be reduced to the
problem of finding a function
$\phi:\sigma\in\mathcal{S}\rightarrow\phi(\sigma)\in\mathbb{R}$ such that
\begin{displaymath}
\phi(\sigma)-3\beta\eta(x)\sum_{y,z\in\Lambda\setminus\{x\}}
                           J_{x,y,z}\sigma(y)\sigma(z)
=
\phi(\eta)-3\beta\sigma(x)\sum_{y,z\in\Lambda\setminus\{x\}}
                           J_{x,y,z}\eta(y)\eta(z) \;\;\; ,
\end{displaymath}
which has no solution.

Let us now discuss the main feature of reversible heat bath
derived probabilistic automata. As it has been seen above, if the
starting Hamiltonian is given by (\ref{h2}), then the Markov chain
(\ref{par}) is reversible with respect to the Gibbs measure with
Hamiltonian $H'$ given by (\ref{ham'}). It is clear that new kind
of interactions, different from the one present in the original
Hamiltonian $H$, arise when $H'$ is considered.

Suppose, for instance, that the starting Hamiltonian has range
$r>0$, namely $J_{x,y}=0$ for any $x,y\in\Lambda$ such that
$|x-y|>r$. Then we have $S_x(\sigma)=\sum_{y:\,0<|x-y|\le
r}J_{x,y}\sigma(y)+h_x$, a sort of average of the spins inside a
ball centered at site $x$ with radius equal to $r$. Hence, by
expanding $H'$ as a sum of potentials we will get all the possible
couplings inside the ball, starting from the two body up to the
$N(r)$ body interaction, with $N(r)$ the number of sites inside
the ball. In some sense these new couplings are dynamically
generated. In the following we will discuss few interesting
particular cases.

\section{Nearest neighbor Ising model}
\label{s:ising}
\par\noindent
Let us consider the standard nearest neighbors Ising model with no
external magnetic field,
namely we consider (\ref{h2}) with $J_{x,y}=J/2$
for any $x,y\in\Lambda$ such that $|x-y|=1$, $J_{x,y}=0$
otherwise, and $h_x=0$ for any $x\in\Lambda$. The Hamiltonian $H'$
is the sum of averages performed over the four site crosses
centered at each site of the lattice. We then expect all the
possible interactions inside the cross.

As it as been seen in \cite{[CN]} it is possible to extract the
potentials and rewrite the Hamiltonian $H'$ in the following way
\begin{equation}
\label{pot}
H'(\sigma)=
-J_1 \sum_{\langle x y
  \rangle_{\!\!\stackrel{\phantom .}{\sqrt{2}}}}
    \sigma(x)\sigma(y)
-J_2 \sum_{\langle x y
  \rangle_{\!\stackrel{\phantom .}{2}}}
    \sigma(x)\sigma(y)
-J_3 \sum_{\diamondsuit_{xywz}}\sigma(x)\sigma(y)\sigma(w)\sigma(z) \;\;\; ,
\end{equation}
where the three sums (see Fig.~1a)
are respectively performed over the pairs of
next to the nearest neighbors (sites at distance $\sqrt{2}$), the
pairs of third neighbors (sites at  distance $2$), the four site
diamond shaped clusters (plaquettes with side length equal to
$\sqrt{2}$). The coupling constants
are given by
\begin{equation}
\label{coup}
J_1=\frac{1}{4}\log\cosh(2\beta J)\sim\frac{1}{2}\beta J,\;\;\;\;\;
J_2=\frac{1}{2} J_1,\;\;\;\;\;
J_3=\frac{1}{16}\log\frac{\cosh^2(2\beta J)}{\cosh^8(\beta J)}
   \sim -\frac{1}{4}\beta J \;\;\; ,
\end{equation}
where ``$\sim$" means the limiting behavior for $\beta\to\infty$.
There exist several possible ways to extract the potentials. A
very natural one, in the case of spin variable, is to consider the
function
$\varphi_x(\sigma)=\log\cosh[(\beta J/2)\sum_{y\in\Delta_x}\sigma(y)]$,
where $\Delta_x=\{y\in\Lambda:\, |y-x|=1\}$ is the set of nearest
neighbors of site $x$,
and its expansion
\begin{displaymath}
\varphi_x(\sigma)=\sum_{X\subset\Delta_x}c(X)\prod_{y\in X}\sigma(y)\;\; ,
\end{displaymath}
with the coefficients $c(X)$ given by
\begin{displaymath}
c(X)=\frac{1}{2^{|\Delta_x|}}\sum_{\sigma\in\{-1,+1\}^{\Delta_x}}
\varphi_x(\sigma) \prod_{y\in X}\sigma(y)\;\; ,
\end{displaymath}
where, we recall, $|\Delta_x|=4$ is the cardinality of $\Delta_x$
(see \cite{[HK]}).

It is important to remark that the second nearest neighbor
interaction, $J_1$, is positive and dominating; hence we expect a
low temperature antiferromagnetic phase to exist. What appears
very remarkable is that we have derived an antiferromagnetic
behavior in a purely dynamical way as a result of the coupling
proliferation. If a parallel heat bath Ising dynamics is
implemented, the equilibrium Gibbs measure shows a low temperature
antiferromagnetic phase despite the simple physical ferromagnetic
coupling of the Ising model. This phenomenon is, obviously, absent if the
Ising heat bath dynamics is implemented in a serial fashion.

The equilibrium properties of the model can be understood by
remarking that two independent models are found if the lattice is
partitioned into two square sublattices with step $\sqrt{2}$ (the
even and the odd sublattice). Each model is, indeed, an eight
vertex model with nearest neighbors coupling $J_1$, next to the
nearest neighbors coupling $J_2$ and plaquette interaction $J_3$.
This model has been widely studied both in two
\cite{[B1],[B2],[NN]} and three \cite{[CCGM],[CG]} dimensions.
{}From (\ref{coup}) and the very well known properties of the
two--dimensional eight vertex model we have that on each
sublattice there are two coexisting low temperature phases,
respectively with positive and negative magnetization. Hence, by
combining in all the possible ways the two phases we get, for the
original model, the $3$ different low temperature phases
corresponding to the three ground states $\psi_0,\psi_1,\psi_2$
(see Fig.~2).

It is of some interest a direct study of the hamiltonian
(\ref{pot}): ground states can be defined as those
configurations on which the Gibbs measure $\mu'$,
associated to the Hamiltonian $H'$, is concentrated when the limit
$\beta\to\infty$ is considered, namely as the minima
of the energy
\begin{equation}
E(\sigma)=
\lim_{\beta\to\infty}\frac{H'(\sigma)}{\beta}
=
-\sum_{x\in\Lambda}|S_x(\sigma)| \;\;\; ,
\label{ener}
\end{equation}
uniformly in $\sigma\in\mathcal{S}$.
It is rather clear that with periodic boundary conditions
there exist three coexisting minima $\psi_0,\psi_1,\psi_2\in\mathcal{S}$
(see Fig.~2),
with energy $-4|\Lambda|$,
such that
$\psi_0(x)=+1$, $\psi_1(x)=(-1)^{x_1+x_2}$ and $\psi_2(x)=-1$
for all $x=(x_1,x_2)\in\Lambda$. Notice that $\psi_1$ is
the chessboard configuration.

The problem is, now, to understand if this coexistence of
different states persists at
a finite small temperature, namely if the system undergoes a low
temperature phase transition.
We give an heuristic argument:
at finite temperature ground states are perturbed because small
droplets of different phases show up. The idea is to calculate the
energetic cost of a perturbation of one of the four
coexisting states via the formation of a square droplet of a
different phase. If it results that one of the three states
$\psi_0, \psi_1,\psi_2$ is more easily perturbed, then we will
conclude that this is the equilibrium phase at finite temperature.

A simple calculation, see \cite{[CN]}, shows that the energy cost
of a square droplet of side length $n$ of one of the two
homogeneous ground states plunged in one of the two chessboards
(or vice versa) is equal to $8n$. On the other hand if an
homogeneous phase is perturbed as above by the other homogeneous
phase, or one of the two chessboards is perturbed by the other
one, then the energy cost is $16n$.

Hence, from the energetical point of view the most convenient
excitations are those in which an homogeneous phase is perturbed
by a chessboard or vice versa. Moreover, for each state
$\psi_0,\psi_1,\psi_2$ there exist two possible energetically
convenient excitations: there is no entropic reason to prefer one
of the four ground states to the others when a finite low
temperature is considered. This remark strongly suggests that at
small finite temperature the three ground states still coexist.

\section{Next to the nearest neighbor Ising model}
\label{s:next}
\par\noindent
Let us consider the Ising model with no external magnetic field
and next to the nearest neighbor interaction, namely we consider
(\ref{h2}) with $J_{x,y}=J/2$ for any $x,y\in\Lambda$ such that
$|x-y|=\sqrt{2}$, $J_{x,y}=0$ otherwise, and $h_x=0$ for any
$x\in\Lambda$.

It is possible to extract the potentials as seen in Section
\ref{s:ising}. The Hamiltonian $H'$ can be written as
\begin{equation}
\label{pot2}
H'(\sigma)=
-J_1 \sum_{\langle x y
  \rangle_{\!\stackrel{\phantom .}{2}}}
    \sigma(x)\sigma(y)
-J_2 \sum_{\langle x y
  \rangle_{\!\stackrel{\phantom .}{2\sqrt{2}}}}
    \sigma(x)\sigma(y)
-J_3 \sum_{\square_{xywz}}\sigma(x)\sigma(y)\sigma(w)\sigma(z) \;\;\; ,
\end{equation}
where the three sums (see Fig.~1b)
are respectively performed over
the pairs of third nearest neighbors (sites at distance $2$),
the pairs of sites at distance $2\sqrt{2}$,
the plaquettes with side length equal to
$2$. The coupling constants are still given by (\ref{coup}).

In order to study this model we remark that if the lattice is
partitioned into four square sublattices with step $2$, then we
obtain four independent models one on each sublattice, each model being
again an eight vertex model with nearest neighbors coupling $J_1$,
next to the nearest neighbors coupling $J_2$ and plaquette interaction $J_3$.
Hence, on each sublattice we have the two degenerate ground states
with all the spins, respectively, equal to one and minus one.
By combining in all the possible ways these two
states we get, for our model, $2^4=16$ different ground states.
On the torus, namely when periodic boundary conditions are
considered, some of the
ground states are equivalent, so we get the seven states
$\psi_0,\psi_1,\dots,\psi_6$ in Fig.~2 and 3.
The fact that the phase transition persists at finite small temperature
is, as in Section \ref{s:ising} a straightforward consequence of the
known behavior of the eight vertex model.

In this note we observed the relation between the two body
ferromagnetic interaction of the Ising model and the low
temperature antiferromagnetic behavior of the equilibrium
Hamiltonian obtained evolving the initial system with a parallel
heat bath dynamics. Any lattice system with two body interactions
and having a self-organization resembling the parallel dynamics
here described in the simple Ising model, could evolve towards
unexpectedly complex equilibrium states. The influence of the
parallel dynamics on the physical properties of the equilibrium
measure emerges as an interesting new feature that deserves
further explorations.

\acknowledgements
\noindent
ENMC wishes to express his thanks to the Physics Department of the
University of Helsinki for its warm hospitality.
ENMC also thanks J.L. Lebowitz for having introduced him to the interesting
topic of PCA's and P. Dai Pra for a useful discussion.
ADP acknowledges support from EU-TMR programme, contract CT98-0169.

\newpage
\begin{center}
\textbf{Figure captions}
\end{center}
\vskip 2 cm
\par\noindent
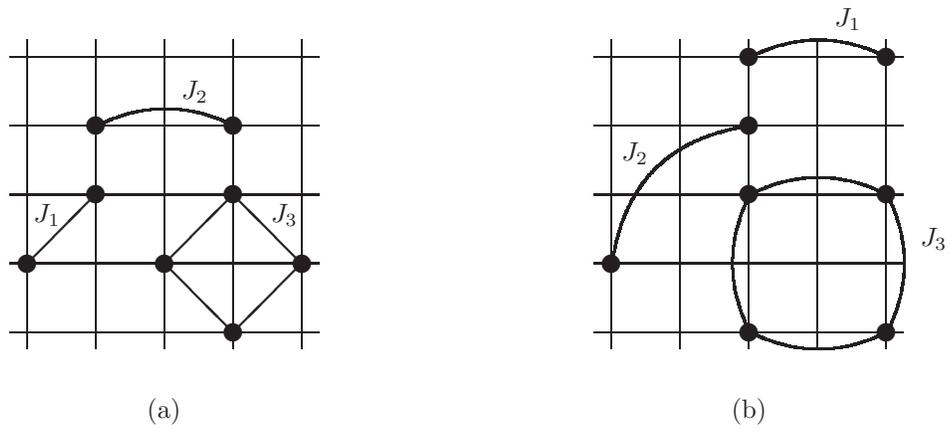
\textbf{Fig.\ 1:}
Couplings $J_1$, $J_2$ and $J_3$  for the hamiltonian (\ref{pot})
(resp.\ (\ref{pot2})) are shown in (a) (resp.\ (b)).
\vskip 0.5 cm
\par\noindent
\textbf{Fig.\ 2:}
The three ground states $\psi_0,\psi_1,\psi_2$ depicted
from the left to the right.
\vskip 0.5 cm
\par\noindent
\textbf{Fig.\ 3:}
The four ground states $\psi_3,\psi_4,\psi_5,\psi_6$
depicted from the left to the right.

\newpage
\setlength{\unitlength}{1.3pt}
\begin{figure}
\begin{picture}(200,200)(-80,-50)
\thinlines
\multiput(-20,0)(20,0){5}{\line(0,1){90}}
\multiput(-25,5)(0,20){5}{\line(1,0){90}}
\thicklines
\put(-20,25){\circle*{5}}
\put(0,45){\circle*{5}}
\put(-20,25){\line(1,1){20}}
\put(-18,37){$J_1$}
\put(0,65){\circle*{5}}
\put(40,65){\circle*{5}}
\qbezier(0,65)(20,75)(40,65)
\put(25,73){$J_2$}
\put(20,25){\circle*{5}}
\put(40,45){\circle*{5}}
\put(40,5){\circle*{5}}
\put(60,25){\circle*{5}}
\put(20,25){\line(1,1){20}}
\put(20,25){\line(1,-1){20}}
\put(60,25){\line(-1,1){20}}
\put(60,25){\line(-1,-1){20}}
\put(51,37){$J_3$}

\put(15,-20){(a)}
\thinlines
\multiput(150,0)(20,0){5}{\line(0,1){90}}
\multiput(145,5)(0,20){5}{\line(1,0){90}}
\thicklines
\put(190,85){\circle*{5}}
\put(230,85){\circle*{5}}
\qbezier(190,85)(210,95)(230,85)
\put(215,94){$J_1$}
\put(150,25){\circle*{5}}
\put(190,65){\circle*{5}}
\qbezier(150,25)(155,60)(190,65)
\put(153,55){$J_2$}
\put(190,45){\circle*{5}}
\put(230,45){\circle*{5}}
\put(190,5){\circle*{5}}
\put(230,5){\circle*{5}}
\qbezier(190,45)(210,55)(230,45)
\qbezier(190,5)(210,-5)(230,5)
\qbezier(190,5)(180,25)(190,45)
\qbezier(230,5)(240,25)(230,45)
\put(240,30){$J_3$}

\put(185,-20){(b)}

\end{picture}
\begin{center}
Fig.\ 1
\end{center}
\label{f:co}
\end{figure}
$\phantom .$

\newpage
\setlength{\unitlength}{1.3pt}
\begin{figure}
\begin{picture}(200,200)(-80,-50)
\thinlines
\put(-20,0){$+$}
\put(0,0){$+$}
\put(-20,20){$+$}
\put(0,20){$+$}

\put(-10,0){$+$}
\put(10,0){$+$}
\put(-10,20){$+$}
\put(10,20){$+$}

\put(-20,10){$+$}
\put(0,10){$+$}
\put(-20,30){$+$}
\put(0,30){$+$}

\put(-10,10){$+$}
\put(10,10){$+$}
\put(-10,30){$+$}
\put(10,30){$+$}

\put(80,0){$+$}
\put(100,0){$+$}
\put(80,20){$+$}
\put(100,20){$+$}

\put(90,0){$-$}
\put(110,0){$-$}
\put(90,20){$-$}
\put(110,20){$-$}

\put(80,10){$-$}
\put(100,10){$-$}
\put(80,30){$-$}
\put(100,30){$-$}

\put(90,10){$+$}
\put(110,10){$+$}
\put(90,30){$+$}
\put(110,30){$+$}

\put(180,0){$-$}
\put(200,0){$-$}
\put(180,20){$-$}
\put(200,20){$-$}

\put(190,0){$-$}
\put(210,0){$-$}
\put(190,20){$-$}
\put(210,20){$-$}

\put(180,10){$-$}
\put(200,10){$-$}
\put(180,30){$-$}
\put(200,30){$-$}

\put(190,10){$-$}
\put(210,10){$-$}
\put(190,30){$-$}
\put(210,30){$-$}

\end{picture}
\begin{center}
Fig.\ 2
\end{center}
\label{f:nn}
\end{figure}
$\phantom .$

\newpage
\setlength{\unitlength}{1.3pt}
\begin{figure}
\begin{picture}(200,200)(-80,-50)
\thinlines
\put(-20,0){$-$}
\put(0,0){$-$}
\put(-20,20){$-$}
\put(0,20){$-$}

\put(-10,0){$+$}
\put(10,0){$+$}
\put(-10,20){$+$}
\put(10,20){$+$}

\put(-20,10){$+$}
\put(0,10){$+$}
\put(-20,30){$+$}
\put(0,30){$+$}

\put(-10,10){$+$}
\put(10,10){$+$}
\put(-10,30){$+$}
\put(10,30){$+$}

\put(40,0){$-$}
\put(60,0){$-$}
\put(40,20){$-$}
\put(60,20){$-$}

\put(50,0){$-$}
\put(70,0){$-$}
\put(50,20){$-$}
\put(70,20){$-$}

\put(40,10){$+$}
\put(60,10){$+$}
\put(40,30){$+$}
\put(60,30){$+$}

\put(50,10){$+$}
\put(70,10){$+$}
\put(50,30){$+$}
\put(70,30){$+$}

\put(100,0){$+$}
\put(120,0){$+$}
\put(100,20){$+$}
\put(120,20){$+$}

\put(110,0){$-$}
\put(130,0){$-$}
\put(110,20){$-$}
\put(130,20){$-$}

\put(100,10){$+$}
\put(120,10){$+$}
\put(100,30){$+$}
\put(120,30){$+$}

\put(110,10){$-$}
\put(130,10){$-$}
\put(110,30){$-$}
\put(130,30){$-$}

\put(160,0){$-$}
\put(180,0){$-$}
\put(160,20){$-$}
\put(180,20){$-$}

\put(170,0){$-$}
\put(190,0){$-$}
\put(170,20){$-$}
\put(190,20){$-$}

\put(160,10){$+$}
\put(180,10){$+$}
\put(160,30){$+$}
\put(180,30){$+$}

\put(170,10){$-$}
\put(190,10){$-$}
\put(170,30){$-$}
\put(190,30){$-$}

\end{picture}
\begin{center}
Fig.\ 3
\end{center}
\label{f:nnn}
\end{figure}
$\phantom .$

\end{document}